\documentstyle [12pt, epsfig]{article}
\begin{document}

\hfill {CUMQ/HEP 123}

\hfill {\today}

\vskip 0.5in   \baselineskip 24pt

{\Large  \bigskip
   \centerline{\Large $B^0_d-{\bar B}^0_d$ mixing and the $B_d 
\rightarrow J/\psi K_s$
   asymmetry}
     \centerline{\Large in the left-right supersymmetric model} }

\vskip .6in
\def\bar{\overline}

\centerline{Mariana Frank \footnote{Email: mfrank@vax2.concordia.ca}
and Shuquan Nie \footnote{Email: sxnie@alcor.concordia.ca}}
\bigskip
\centerline {\it Department of Physics, Concordia University, 1455 De
Maisonneuve Blvd. W.}
\centerline {\it Montreal, Quebec, Canada, H3G 1M8}

\vskip 0.5in

{\narrower\narrower We analyze $B^0_d-{\bar B}^0_d$ mixing and the 
$B_d \rightarrow J/\psi K_s$
   asymmetry in a fully left-right supersymmetric model. We give 
explicit expressions
for all the chargino, gluino, gluino-neutralino and neutralino 
amplitudes involved in
$B^0_d-{\bar B}^0_d$ mixing. We calculate the mass difference $\Delta 
m_{B_d}$  and the
CP asymmetry $a_{J/\psi K_s}$, allowing for supersymmetric sources of 
flavor violation
arising from squark mixings. We obtain conservative constraints on 
the various squark
mass splittings and compare these with analyses performed in other models.}

   PACS number(s): 12.60.Jv, 13.25.Hw, 14.40.Nd

\newpage

\section{Introduction}

Flavor changing neutral currents (FCNC) and charge parity (CP) violating
phenomena are some of the best probes for physics beyond the Standard
Model (SM). All existing measurements so far are consistent with the
SM predictions involving the Cabibbo-Kobayashi-Maskawa (CKM) matrix as the
only source of flavor violation.

In the SM, FCNC are absent at tree level,
appear at one loop level, but they are effectively suppressed by the
Glashow-Iliopulos-Maiani (GIM) mechanism and
small CKM angles. In supersymmetric models, there is no similar mechanism to
suppress the loop contributions to either flavor or CP violating
phenomena. Experimental studies of flavor physics, especially in B
decays, appear essential for the understanding of the mechanism for
supersymmetry breaking. With the increased statistical power of
experiments at B factories, rare B decays will be measured very
precisely.

CP violation arises in the SM from complex couplings
in the charged current, leading to a physical phase in the
CKM matrix. In supersymmetric extensions of
the SM there are additional sources of CP violation, due to the
presence of new phases in the supersymmetric Lagrangian. CP violation was
observed first in the kaon system \cite{kaon}. Recently both BaBar 
\cite{babar} and
BELLE \cite{belle} collaborations have provided clear evidence for CP
violation in the B-system, although at present the experimental errors are
relatively large. Even if these observed CP asymmetries roughly 
agree, within errors,
with the SM prediction, there is still considerable space
available for new physics, and supersymmetry particularly. But new
phases introduced by supersymmetry must be constrained by
electric dipole moments of the neutron, electron, and Mercury
atom \cite{EDM}.

The newly measured CP asymmetry, $a_{J/\Psi K_s}$, in the decay
$B \rightarrow J/\Psi K_s$ is
defined as follows
\begin{equation}
a_{J/\Psi K_s}(t)= \frac{\Gamma (B_d (t) \rightarrow J/\Psi K_s)-\Gamma
({\bar B}_d (t) \rightarrow J/\Psi K_s)}
{\Gamma (B_d (t) \rightarrow J/\Psi K_s)+\Gamma
({\bar B}_d (t) \rightarrow J/\Psi K_s)} =-a_{J/\Psi K_s} \sin (\Delta
m_{B_d}t).
\end{equation}
BaBar and BELLE have announced the following results
\begin{eqnarray}
a_{J/\Psi K_s}& = &0.59 \pm 0.14 \pm 0.05~~~{\rm (BaBar)},\nonumber \\
a_{J/\Psi K_s}& = &0.99 \pm 0.14 \pm 0.06~~~{\rm (BELLE)}.
\end{eqnarray}
The present world average is $a_{J/\Psi K_s} = 0.79 \pm 0.12$
\cite{worldav}.
In the SM, $a_{J/\Psi K_s}$ is related to the inner angle of
the unitarity triangle
\begin{equation}
a_{J/\Psi K_s}^{SM}= \sin 2\beta~;~~~\beta= arg (-\frac
{V_{cd}V_{cb}^{\ast}}{V_{td}V_{tb}^{\ast}}).
\end{equation}

Rare B decays involving loop induced flavor changing neutral transitions are
sensitive to the properties of the internal heavy particles, making 
them particularly
suitable as probes of physics beyond the SM. It can be expected that there
are considerable new physics contributions to $B^0_d-{\bar B}^0_d$ mixing.
However, these contributions appear small in a supergravity-inspired MSSM
\cite{bertolini}. Therefore $B^0_d-{\bar B}^0_d$ provides for excellent
opportunities to test physics beyond MSSM.

$\Delta B=2$ decays have been studied
in the framework of supersymmetric models with universal soft
supersymmetry breaking terms
\cite{b=2}. It was shown that non-universal realizations of SUSY could
give large contributions to $\Delta F=2$ observables \cite{piai}, 
making them distinguishable
from the MSSM. It becomes possible in this scenario to discover SUSY indirectly
in precision measurements  of B-physics.

Although some attempts have been
made to reconcile $\Delta B=2$ with right-handed $b$-quark
decays \cite{ch}, a complete analysis of the $B^0_d-{\bar B}^0_d$ mixing
in a fully left-right supersymmetric model is still lacking.
In our previous work \cite{fn, bsl1l2}, we
analyzed the $\Delta B=1$ processes in the context of the left-right
supersymmetric model and found new contributions. We also found that
these processes place tight bounds on supersymmetric flavor violation
parameters. We extend this work here to $\Delta B=2$ processes with the hope of
adding one more piece to the puzzle of B physics.

The Left-Right Supersymmetric (LRSUSY) models \cite{history, frank1}, based
on the group $SU(2)_L
\times SU(2)_R \times U(1)_{B-L}$, incorporate the advantage of supersymmetry
within a natural framework for allowing neutrino masses through the
seesaw mechanism
\cite{mohapatra}. LRSUSY models can be embedded in a supersymmetric 
grand unified
theory such as
$SO(10)$ \cite{SO10}. They would also appear in building realistic brane
worlds from Type I strings. This involves left-right
supersymmetry, with supersymmetry broken either at the string scale
$M_{SUSY} \approx 10^{10-12}$ GeV, or at $M_{SUSY} \approx 1$ TeV, the
difference having implications for the gauge unification \cite{string}.

In this paper we study all contributions of the LRSUSY model to
the $B_d^0-{\bar B}^0_d$ mixing at one-loop level. The process can be
mediated not only by left- and right-handed W bosons and charged Higgs
bosons as in the nonsupersymmetric case, but also by
charginos, neutralinos and gluinos. The structure of the LRSUSY model
provides significant contributions from the right-handed squarks
and an enlarged gaugino-higgsino sector with right-handed couplings,
which are not as constrained as the
right-handed gauge sector in left-right symmetric models. We anticipate
that these would contribute a large enhancement of the mass difference
and CP asymmetry
and would constrain the parameter space of the model.

The paper is organized as follows. In Sec. II, we review the main features
of LRSUSY and give the supersymmetric contributions to the $\Delta
B=2$ process.  In Sec. III,
we present the numerical analysis and conservative bounds on various 
squark mass
splittings are obtained. We reach our conclusions in Sec. IV.

\section{The analytic formulas}

The minimal supersymmetric left-right model is based on the gauge group
$SU(3)_C \times SU(2)_L \times SU(2)_R \times U(1)_{B-L}$. The matter
fields of this model consist of three families of quark and lepton chiral
superfields transforming as the adjoint representations of the groups. The
Higgs sector consists of the bidoublet and triplet Higgs superfields:
\begin{eqnarray}
\Phi_1 = \left (\begin{array}{cc}
\Phi^0_{11}&\Phi^+_{11}\\ \Phi_{12}^-& \Phi_{12}^0
\end{array}\right),~~~
\Phi_2=\left (\begin{array}{cc}
\Phi^0_{21}&\Phi^+_{21}\\ \Phi_{22}^-& \Phi_{22}^0
\end{array}\right) \nonumber \\
\Delta_{L}  = \left(\begin{array}{cc}
\frac {1}{\sqrt{2}}\Delta_L^-&\Delta_L^0\\
\Delta_{L}^{--}&-\frac{1}{\sqrt{2}}\Delta_L^-
\end{array}\right) ,~~~\delta_{L}  =
\left(\begin{array}{cc}
\frac {1}{\sqrt{2}}\delta_L^+&\delta_L^{++}\\
\delta_{L}^{0}&-\frac{1}{\sqrt{2}}\delta_L^+
\end{array}\right) \nonumber \\
\Delta_{R}  =
\left(\begin{array}{cc}
\frac {1}{\sqrt{2}}\Delta_R^-&\Delta_R^0\\
\Delta_{R}^{--}&-\frac{1}{\sqrt{2}}\Delta_R^-
\end{array}\right),~~~\delta_{R}  =
\left(\begin{array}{cc}
\frac {1}{\sqrt{2}}\delta_R^+&\delta_R^{++}\\
\delta_{R}^{0}&-\frac{1}{\sqrt{2}}\delta_R^+
\end{array}\right).
\end{eqnarray}
The bidoublet Higgs superfields appear in all LRSUSY and serve to
implement the $SU(2)_L \times U(1)_{Y}$ symmetry breaking and to generate
the CKM mixing matrix. Supplementary Higgs
representations are needed to break left-right symmetry spontaneously:
either doublets or triplets would achieve this, but the triplet Higgs
$\Delta_L, ~\Delta_R$ bosons have the advantage of supporting the seesaw
mechanism. Since the theory is supersymmetric, additional
triplet superfields $\delta_L, ~\delta_R$ are needed to cancel triangle
gauge anomalies in the fermionic sector. The symmetry is broken 
spontaneously to
$U(1)_{em}$.
There are three different stages of symmetry breakdown. At the first
stage only discrete parity is broken. In the second stage of
symmetry breaking, LRSUSY is broken down to the MSSM at $\Lambda_R$ 
by the vevs of the
neutral triplet Higgs bosons $\langle \Delta_R \rangle \neq
0,~~\langle \delta_R \rangle \neq 0$.
  The final stage of symmetry breakdown takes place
at electroweak scales $\Lambda_L$ and MSSM is broken down to $U(1)_{em}$
through bidoublet vevs $\kappa_{1},~\kappa_{2} \neq 0$.
In addition, supersymmetry can be broken at any scale between $\Lambda_R$ and
$\Lambda_L$.

The most general superpotential involving these superfields is
\begin{eqnarray}
\label{superpotential}
W & = & {\bf Y}_{Q}^{(i)} Q^T\Phi_{i}i \tau_{2}Q^{c} + {\bf Y}_{L}^{(i)}
L^T \Phi_{i}i \tau_{2}L^{c} + i{\bf Y}_{LR}(L^T\tau_{2} \delta_L L +
L^{cT}\tau_{2}
\Delta_R L^{c}) \nonumber \\
& & + \mu_{LR}\left [Tr (\Delta_L  \delta_L +\Delta_R
\delta_R)\right] + \mu_{ij}Tr(i\tau_{2}\Phi^{T}_{i} i\tau_{2} \Phi_{j})
+W_{NR}
\end{eqnarray}
where $W_{NR}$ denotes (possible) non-renormalizable terms arising 
from higher scale
physics or Planck scale effects~\cite{recmohapatra}. The presence of 
these terms
insures that, when the SUSY breaking scale is above $\Lambda_{R}$, the
ground state is R-parity conserving ~\cite{km}. In addition, the
potential also includes well-known $F$-terms, $D$-terms as well as soft
supersymmetry breaking terms:
\begin{eqnarray}
\label{eq:soft}
{\cal L}_{soft}&=&\left[ {\bf A}_{Q}^{i}{\bf Y}_{Q}^{(i)}{\tilde Q}^T\Phi_{i}
i\tau_{2}{\tilde Q}^{c}+ {\bf A}_{L}^{i}{\bf Y}_{L}^{(i)}{\tilde L}^T \Phi_{i}
i\tau_{2}{\tilde L}^{c} + i{\bf A}_{LR} {\bf Y}_{LR}({\tilde L}^T\tau_{2}
\delta_L{\tilde  L} + L^{cT}\tau_{2} \Delta_R{\tilde L}^{c})
\right.
\nonumber
\\ & & \left. +({ m}_{\Phi}^{ 2})_{ij}
\Phi_i^{\dagger}  \Phi_j \right] + \left[( m_{L}^2)_{ij}{\tilde 
l}_{Li}^{\dagger}{\tilde
l}_{Lj}+ (m_{R}^2)_{ij}{\tilde l}_{Ri}^{\dagger}{\tilde l}_{Rj} \right]
\nonumber \\
& & - M_{LR}^2 \left[Tr(  \Delta_R  \delta_R)+ Tr(  \Delta_L
  \delta_L) +h.c.\right] - [B \mu_{ij} \Phi_{i} \Phi_{j}+h.c.]
\end{eqnarray}
These parts of the Lagrangian are responsible for flavor violation in 
lepton and quark
decays in general, and in the B system in particular.

\begin{figure}
\centerline{ \epsfysize 2.0in
\rotatebox{360}{\epsfbox{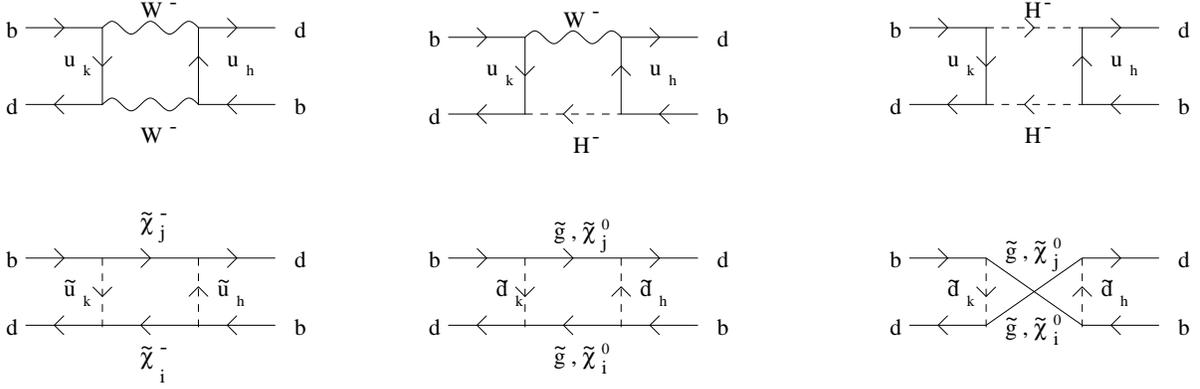}}  }
\caption{ Leading box diagrams contributing to $B^0_d-\bar{B}^0_d$
mixing. }
\protect \label{feynman}
\end{figure}

The contributions of the left-right supersymmetric model to the  $B^0_q
-{\bar B}^0_q ~(q=d,~s)$ mixing are given by the effective Hamiltonian
\begin{equation}
{\cal H}_{eff}^{\Delta B=2} =\sum_i [ C_i(\mu) Q_i(\mu) +
{\tilde C}_i(\mu) {\tilde Q}_i(\mu)].
\end{equation}
where the relevant operators entering the sum are
\begin{eqnarray}
Q_{1}&=&{\bar q}_L^\alpha \gamma_\mu b_L^\alpha {\bar
q}_L^\beta \gamma_\mu b_L^\beta,
\nonumber \\
{\tilde Q}_{1}&=&{\bar q}_R^\alpha \gamma_\mu b_R^\alpha {\bar
q}_R^\beta \gamma_\mu b_R^\beta,
\nonumber \\
Q_{2}&=&{\bar q}_L^\alpha b_R^\alpha {\bar
q}_L^\beta b_R^\beta,
\nonumber \\
{\tilde Q}_{2}&=&{\bar q}_R^\alpha b_L^\alpha {\bar
q}_R^\beta b_L^\beta,
\nonumber \\
Q_{3}&=&{\bar q}_L^\alpha  b_R^\beta {\bar
q}_L^\beta  b_R^\alpha,
\nonumber \\
{\tilde Q}_3&=&{\bar q}_R^\alpha  b_L^\beta {\bar
q}_R^\beta  b_L^\alpha,
\nonumber \\
Q_{4}&=&{\bar q}_L^\alpha b_R^\alpha {\bar
q}_R^\beta b_L^\beta,
\nonumber \\
Q_{5}&=&{\bar q}_L^\alpha  b_R^\beta {\bar
q}_R^\beta  b_L^\alpha.
\end{eqnarray}
The Wilson coefficients $C_{i}$ and ${\tilde C}_i$ are initially evaluated
at the electroweak or soft supersymmetry breaking scale, then evolved down
to the scale $\mu$. In the SM and constrained SUSY models, ${\tilde
Q}_i$ contributions
are generally suppressed by ${\cal O}(m_q/m_b)$ compared with the
contributions from $Q_i$.
However this is not the case in generic SUSY models such as
non-universal models, or in left-right models. Because of left-right symmetry,
we must consider all contributions from both chirality operators.
The $B^0_d-{\bar B}^0_d$ mixing is mediated through the box diagrams 
in Fig. \ref{feynman}.

The SM and left-right symmetric contributions to the
$\Delta B=2$ transitions have been discussed before
and are well-known \cite{SMLR}. In general, the SM contribution will 
be added to any
new contributions and their sum would have to saturate the 
experimental value $\Delta
m_{B_d}=0.489$ /ps within 10\% errors.  However, since this sum 
depends on the relative
phase of the two contributions, thus introducing a new parameter, we adopt a
simplifying assumption. The most common assumption in the literature is to set
the SM contribution to the mass difference $\Delta m_{B_d}$ to zero, 
then require that
the supersymmetric contribution of each independent combination of mass
insertions does not exceed the central value $\Delta m_{B_d} 
<0.489~(ps)^{-1}$. This
would set
  conservative upper bounds on various mass insertions
\cite{b=2}. An alternative assumption imposes the sum of the SUSY 
contributions and
the SM contributions not to exceed the experimentally measured values 
of $\Delta m_{B_d}$
and $\sin 2 \beta$ by more than $1
\sigma$ \cite{becirevic}. We adopt the former ansatz. Below we give a 
comprehensive list
of all box diagram contributions. We refer to Ref. \cite{ fn, bsl1l2} for the
definitions of the notations and functions used.

\subsection{The Chargino Contribution}
\begin{eqnarray}
C_1^{\tilde{\chi}^-} &=& \frac{\alpha_W^2}{16} \sum_{h,k=1}^6 \sum_{i,j=1}^{5}
\frac{1}{m_{\tilde{\chi}_j^-}^2}
(G_{UL}^{jkb}-H_{UR}^{jkb}) (G_{UL}^{\star ikq}-H_{UR}^{\star ikq}) 
\nonumber \\
& & (G_{UL}^{ihb}-H_{UR}^{ihb})(G_{UL}^{\star jhq}-H_{UR}^{\star jhq})
G^{\prime} (x_{\tilde{u}_k \tilde{\chi}_j^-},
x_{\tilde{u}_h \tilde{\chi}_j^-},x_{\tilde{\chi}_i \tilde{\chi}_j^-}),
\nonumber \\
\tilde{C}_1^{\tilde{\chi}^-} &=& C_1^{\tilde{\chi}^-} (L \leftrightarrow R),
\nonumber \\
C_3^{\tilde{\chi}^-} &=& -\frac{\alpha_W^2}{4} \sum_{h,k=1}^6 \sum_{i,j=1}^{5}
\frac{1}{m_{\tilde{\chi}_j^-}^2}
(G_{UR}^{jkb}-H_{UL}^{jkb}) (G_{UL}^{\star ikq}-H_{UR}^{\star ikq}) 
\nonumber \\
& & (G_{UR}^{ihb}-H_{UL}^{ihb})(G_{UL}^{\star jhq}-H_{UR}^{\star jhq})
\sqrt{x_{\tilde{\chi}_i^- \tilde{\chi}_j^-}} F^{\prime} 
(x_{\tilde{u}_k \tilde{\chi}_j^-},
x_{\tilde{u}_h \tilde{\chi}_j^-},x_{\tilde{\chi}_i \tilde{\chi}_j^-}),
\nonumber \\
\tilde{C}_3^{\tilde{\chi}^-} &=& C_3^{\tilde{\chi}^-} (L \leftrightarrow R),
\nonumber \\
C_4^{\tilde{\chi}^-} &=& \frac{\alpha_W^2}{4} \sum_{i,j=1}^5 
\sum_{h,k=1}^6 \frac{1}{m_{\tilde{\chi}_j^-}^2}
(2 H_{UR}^{jkb} H_{UL}^{ihb} G_{UL}^{\star ikq} G_{UR}^{\star jhq}
+2 H_{UL}^{jkb} H_{UR}^{jhb} G_{UL}^{\star ihq} G_{UR}^{\star ikq}
\nonumber \\
& & - G_{UR}^{jkb} G_{UL}^{ihb} G_{UL}^{\star jkq} G_{UR}^{\star ihq})
  G^{\prime} (x_{\tilde{u}_k \tilde{\chi}_j^-}, x_{\tilde{u}_h 
\tilde{\chi}_j^-},
x_{\tilde{\chi}_i \tilde{\chi}_j^-}),
\nonumber \\
C_5^{\tilde{\chi}^-} &=& \frac{\alpha_W^2}{4} \sum_{i,j=1}^5 
\sum_{h,k=1}^6 \frac{1}{m_{\tilde{\chi}_j^-}^2}
[(4 G_{UR}^{jkb} G_{UL}^{jhb} G_{UR}^{\star ikq} G_{UL}^{\star ihq}
- G_{UL}^{jkb} G_{UR}^{jhb} G_{UL}^{\star ihq} G_{UR}^{\star ikq})
\nonumber \\
& &  G^{\prime} (x_{\tilde{u}_k \tilde{\chi}_j^-}, x_{\tilde{u}_h 
\tilde{\chi}_j^-},
x_{\tilde{\chi}_i \tilde{\chi}_j^-}) -(4 G_{UR}^{jkb} G_{UL}^{ihb} 
G_{UR}^{\star ihq} G_{UL}^{\star jkq}
- G_{UL}^{jkb} G_{UR}^{ihb} G_{UL}^{\star ihq} G_{UR}^{\star jkq})
\nonumber \\
& & 2 \sqrt{x_{\tilde{\chi}_i^- \tilde{\chi}_j^-}}
F^{\prime} (x_{\tilde{u}_k \tilde{\chi}_j^-}, x_{\tilde{u}_h \tilde{\chi}_j^-},
x_{\tilde{\chi}_i \tilde{\chi}_j^-})] .
\end{eqnarray}
There are no chargino contributions to $C_2$ and $\tilde{C}_2$ because of the
color structure of the chargino box diagram.

\subsection{The Gluino Contribution}
\begin{eqnarray}
C_1^{\tilde{g}}&=& \frac{\alpha_s^2}{2 m^2_{\tilde{g}}} \sum_{h,k=1}^{6}
\Gamma_{DL}^{kb} \Gamma_{DL}^{\star kq} \Gamma_{DL}^{hb} \Gamma_{DL}^{\star hq}
[\frac{11}{9} G(x_{\tilde{d}_h \tilde{g}}, x_{\tilde{d}_k \tilde{g}})
-\frac{1}{9} F(x_{\tilde{d}_h \tilde{g}}, x_{\tilde{d}_k \tilde{g}})],
\nonumber \\
{\tilde C}_1^{\tilde{g}}&=& C_1^{\tilde{g}} ( L \leftrightarrow R ) ,
\nonumber \\
C_2^{\tilde{g}} &=& - \frac{\alpha_s^2}{2 m^2_{\tilde{g}}} \sum_{h,k=1}^{6}
\Gamma_{DR}^{kb} \Gamma_{DL}^{\star kq} \Gamma_{DR}^{hb} \Gamma_{DL}^{\star hq}
\frac{17}{18} F(x_{\tilde{d}_h \tilde{g}}, x_{\tilde{d}_k \tilde{g}}),
\nonumber \\
\tilde{C}_2^{\tilde{g}} &=&  C_2^{\tilde{g}}( L \leftrightarrow R )   ,
\nonumber \\
C_3^{\tilde{g}} &=& \frac{\alpha_s^2}{2 m^2_{\tilde{g}}} \sum_{h,k=1}^{6}
\Gamma_{DR}^{kb} \Gamma_{DL}^{\star hq} \Gamma_{DR}^{hb} \Gamma_{DL}^{\star kq}
\frac{1}{6} F(x_{\tilde{d}_h \tilde{g}}, x_{\tilde{d}_k \tilde{g}}),
\nonumber \\
\tilde{C}_3^{\tilde{g}} &=& C_3^{\tilde{g}} ( L \leftrightarrow R )  ,
\nonumber \\
C_4^{\tilde{g}}&=& -\frac{\alpha_s^2}{2 m^2_{\tilde{g}}} \sum_{h,k=1}^{6}
\{ \Gamma_{DR}^{kb} \Gamma_{DL}^{\star kq} \Gamma_{DL}^{hb} 
\Gamma_{DR}^{\star hq}
[\frac{1}{3} G(x_{\tilde{d}_h \tilde{g}}, x_{\tilde{d}_k \tilde{g}})
+\frac{7}{3} F(x_{\tilde{d}_h \tilde{g}}, x_{\tilde{d}_k \tilde{g}})]
\nonumber \\
& & + \Gamma_{DR}^{kb} \Gamma_{DL}^{\star kq} \Gamma_{DL}^{hb} 
\Gamma_{DR}^{\star hq}
\frac{11}{18} G(x_{\tilde{d}_h \tilde{g}}, x_{\tilde{d}_k \tilde{g}}) \},
\nonumber \\
C_5^{\tilde{g}}&=& \frac{\alpha_s^2}{2 m^2_{\tilde{g}}} \sum_{h,k=1}^{6}
\{ \Gamma_{DL}^{kb} \Gamma_{DL}^{\star kq} \Gamma_{DR}^{hb} 
\Gamma_{DR}^{\star hq}
[\frac{5}{9} G(x_{\tilde{d}_h \tilde{g}}, x_{\tilde{d}_k \tilde{g}})
-\frac{1}{9} F(x_{\tilde{d}_h \tilde{g}}, x_{\tilde{d}_k \tilde{g}})]
\nonumber \\
& & - \Gamma_{DR}^{kb} \Gamma_{DL}^{\star kq} \Gamma_{DL}^{hb} 
\Gamma_{DR}^{\star hq}
\frac{5}{6} G(x_{\tilde{d}_h \tilde{g}}, x_{\tilde{d}_k \tilde{g}}) \}.
\end{eqnarray}
These terms include the box and the crossed diagrams.

\subsection{The Neutralino Contribution}
\begin{eqnarray}
C_1^{\tilde{\chi}^0}&=& \frac{\alpha_W^2}{4} \sum_{h,k=1}^{6} \sum_{i,j=1}^9
\frac{1}{m_{\tilde{\chi}_j^0}^2}
[G_{0DL}^{jkb} G_{0DL}^{\star ikq} G_{0DL}^{ihb} G_{0DL}^{\star jhq}
G^{\prime}(x_{\tilde{d}_k \tilde{\chi}_j^0},
x_{\tilde{d}_h \tilde{\chi}_j^0},
x_{\tilde{\chi}_i^0 \tilde{\chi}_j^0})
\nonumber \\
& &  -G_{0DL}^{jkb} G_{0DL}^{\star ikq} G_{0DL}^{jhb} G_{0DL}^{\star ihq}
2 \sqrt{x_{\tilde{\chi}_i^0 \tilde{\chi}_j^0}}
  F^{\prime}(x_{\tilde{d}_k \tilde{\chi}_j^0}, x_{\tilde{d}_h \tilde{\chi}_j^0},
x_{\tilde{\chi}_i^0 \tilde{\chi}_j^0})],
\nonumber \\
{\tilde C}_1^{\tilde{\chi}^0}&=&  C_1^{\tilde{\chi}^0} (L \leftrightarrow R),
\nonumber \\
C_2^{\tilde{\chi}^0}&=& \frac{\alpha_W^2}{2} \sum_{h,k=1}^{6} \sum_{i,j=1}^9
\frac{1}{m_{\tilde{\chi}_j^0}^2}
(H_{0DL}^{jkb} H_{0DL}^{jhb} G_{0DL}^{\star ikq} G_{0DL}^{\star ihq}
-G_{0DR}^{jkb} G_{0DR}^{jhb} G_{0DL}^{\star ikq} G_{0DL}^{\star ihq})
\nonumber \\
& & \sqrt{x_{\tilde{\chi}_i^0 \tilde{\chi}_j^0}}
  F^{\prime}(x_{\tilde{d}_k \tilde{\chi}_j^0}, x_{\tilde{d}_h \tilde{\chi}_j^0},
x_{\tilde{\chi}_i^0 \tilde{\chi}_j^0}),
\nonumber \\
\tilde{C}_2^{\tilde{\chi}^0}&=&C_2^{\tilde{\chi}^0}(L \leftrightarrow R),
\nonumber \\
C_3^{\tilde{\chi}^0}&=& \frac{\alpha_W^2}{2} \sum_{h,k=1}^{6} \sum_{i,j=1}^9
\frac{1}{m_{\tilde{\chi}_j^0}^2}
\sqrt{x_{\tilde{\chi}_i^0 \tilde{\chi}_j^0}}
  F^{\prime}(x_{\tilde{d}_k \tilde{\chi}_j^0}, x_{\tilde{d}_h \tilde{\chi}_j^0},
x_{\tilde{\chi}_i^0 \tilde{\chi}_j^0})
\nonumber \\
& &  \times [H_{0DL}^{jkb} H_{0DL}^{jhb} G_{0DL}^{\star ikq} 
G_{0DL}^{\star ihq}
- H_{0DL}^{jkb} H_{0DL}^{ihb} G_{0DL}^{\star jkq} G_{0DL}^{\star ihq}
\nonumber \\
& & + G_{0DR}^{jkb} G_{0DR}^{jhb} G_{0DL}^{\star ikq} G_{0DL}^{\star ihq}
- G_{0DR}^{jkb} G_{0DR}^{ihb} G_{0DL}^{\star jkq} G_{0DL}^{\star ihq}],
\nonumber \\
\tilde{C}_3^{\tilde{\chi}^0}&=& C_3^{\tilde{\chi}^0} (L \leftrightarrow R),
\nonumber \\
C_4^{\tilde{\chi}^0}&=& \frac{\alpha_W^2}{4} \sum_{h,k=1}^{6} \sum_{i,j=1}^9
\frac{1}{m_{\tilde{\chi}_j^0}^2}
G^{\prime}(x_{\tilde{d}_k \tilde{\chi}_j^0}, x_{\tilde{d}_h \tilde{\chi}_j^0},
x_{\tilde{\chi}_i^0 \tilde{\chi}_j^0}) [ 2 H_{0DR}^{jkb} 
H_{0DL}^{ihb} G_{0DL}^{\star ikq} G_{0DR}^{\star jhq}
\nonumber \\
& &
+ 2 H_{0DL}^{jkb} H_{0DR}^{jhb} G_{0DL}^{\star ikq} G_{0DR}^{\star ihq}
-G_{0DR}^{jkb} G_{0DL}^{\star jkq} G_{0DL}^{ihb} G_{0DR}^{\star ihq} ],
\nonumber \\
C_5^{\tilde{\chi}^0}&=& \frac{\alpha_W^2}{4} \sum_{h,k=1}^{6} \sum_{i,j=1}^9
\frac{1}{m_{\tilde{\chi}_j^0}^2}
[(4 G_{0DR}^{jkb} G_{0DL}^{jhb} G_{0DR}^{\star ikq} G_{0DL}^{\star ihq}
-G_{0DL}^{jkb} G_{0DR}^{jhb} G_{0DR}^{\star ikq} G_{0DL}^{\star ihq})
\nonumber \\
& & G^{\prime}(x_{\tilde{d}_k \tilde{\chi}_j^0},
x_{\tilde{d}_h \tilde{\chi}_j^0},
x_{\tilde{\chi}_i^0 \tilde{\chi}_j^0})
-(4 G_{0DR}^{jkb} G_{0DL}^{\star jkq} G_{0DL}^{ihb} G_{0DR}^{\star ihq}
-G_{0DL}^{jkb} G_{0DR}^{ihb} G_{0DL}^{\star ihq} G_{0DR}^{\star jkq})
\nonumber \\
& & 2 \sqrt{x_{\tilde{\chi}_i^0 \tilde{\chi}_j^0}}
  F^{\prime}(x_{\tilde{d}_k \tilde{\chi}_j^0}, x_{\tilde{d}_h \tilde{\chi}_j^0},
x_{\tilde{\chi}_i^0 \tilde{\chi}_j^0})].
\end{eqnarray}

\subsection{The Gluino-Neutralino Contribution}
\begin{eqnarray}
C_1^{\tilde g \tilde{\chi}^0}&=& \frac{\alpha_s \alpha_W}{6 m_{\tilde g}^2}
\sum_{h,k=1}^{6} \sum_{i=1}^9
[\Gamma_{DL}^{\star kq} \Gamma_{DL}^{hb} G_{0DL}^{ikb} G_{0DL}^{\star ihq}
G^{\prime}(x_{\tilde{d}_k \tilde{g}},
x_{\tilde{d}_h \tilde{g}},
x_{\tilde{\chi}_i^0 \tilde{g}})
\nonumber \\
& &  -(\Gamma_{DL}^{kb} \Gamma_{DL}^{hb} G_{0DL}^{\star ikq} 
G_{0DL}^{\star ihq}
+\Gamma_{DL}^{\star kq} \Gamma_{DL}^{\star hq} G_{0DL}^{ikb} G_{0DL}^{ihb})
\sqrt{x_{\tilde{\chi}_i^0 \tilde{g}}}
F^{\prime}(x_{\tilde{d}_k \tilde{g}}, x_{\tilde{d}_h \tilde{g}},
x_{\tilde{\chi}_i^0 \tilde{g}})],
\nonumber \\
\tilde C_1^{\tilde g \tilde{\chi}^0}&=& C_1^{\tilde g 
\tilde{\chi}^0}(L \leftrightarrow R),
\nonumber \\
C_2^{\tilde g \tilde{\chi}^0}&=& \frac{\alpha_s \alpha_W}{4 m_{\tilde g}^2}
\sum_{h,k=1}^{6} \sum_{i=1}^9
  (\frac{2}{3} \Gamma_{DL}^{\star kq} \Gamma_{DL}^{\star hq} 
H_{0DL}^{ikb} H_{0DL}^{ihb}
-\Gamma_{DR}^{kb} \Gamma_{DR}^{hb} G_{0DL}^{\star ikq} G_{0DL}^{\star ihq})
\sqrt{x_{\tilde{\chi}_i^0 \tilde{g}}}
F^{\prime}(x_{\tilde{d}_k \tilde{g}}, x_{\tilde{d}_h \tilde{g}},
x_{\tilde{\chi}_i^0 \tilde{g}}),
\nonumber \\
\tilde C_2^{\tilde g \tilde{\chi}^0}&=& C_2^{\tilde g 
\tilde{\chi}^0}(L \leftrightarrow R),
\nonumber \\
C_3^{\tilde g \tilde{\chi}^0}&=& \frac{\alpha_s \alpha_W}{4 m_{\tilde g}^2}
\sum_{h,k=1}^{6} \sum_{i=1}^9
  (\frac{2}{3} \Gamma_{DL}^{\star kq} \Gamma_{DL}^{\star hq} 
H_{0DL}^{ikb} H_{0DL}^{ihb}
-\Gamma_{DR}^{kb} \Gamma_{DR}^{hb} G_{0DL}^{\star ikq} G_{0DL}^{\star ihq})
\sqrt{x_{\tilde{\chi}_i^0 \tilde{g}}}
F^{\prime}(x_{\tilde{d}_k \tilde{g}}, x_{\tilde{d}_h \tilde{g}},
x_{\tilde{\chi}_i^0 \tilde{g}}),
\nonumber \\
\tilde C_3^{\tilde g \tilde{\chi}^0}&=& C_3^{\tilde g 
\tilde{\chi}^0}(L \leftrightarrow R),
\nonumber \\
C_4^{\tilde g \tilde{\chi}^0}&=& \frac{\alpha_s \alpha_W}{2 m_{\tilde g}^2}
\sum_{h,k=1}^{6} \sum_{i=1}^9 [
( \Gamma_{DR}^{\star kq} \Gamma_{DL}^{\star hq} G_{0DR}^{ikb} G_{0DL}^{ihb}
-\frac{1}{3} \Gamma_{DR}^{ \star kq} \Gamma_{DL}^{\star hq} 
H_{0DR}^{ikb} H_{0DL}^{ihb}
+\Gamma_{DR}^{kb} \Gamma_{DL}^{hb} G_{0DR}^{\star ikq} G_{0DL}^{\star ihq}
\nonumber \\
& & - \frac{1}{3} \Gamma_{DR}^{kb} \Gamma_{DR}^{ \star hq} 
G_{0DL}^{ihb} G_{0DL}^{\star ikq}
- \frac{1}{3} \Gamma_{DL}^{hb} \Gamma_{DL}^{ \star kq} G_{0DR}^{ikb} 
G_{0DR}^{\star ihq})
G^{\prime}(x_{\tilde{d}_k \tilde{g}},
x_{\tilde{d}_h \tilde{g}},
x_{\tilde{\chi}_i^0 \tilde{g}}  )
\nonumber \\
& &  -2  ( \Gamma_{DL}^{\star kq} \Gamma_{DR}^{hb} G_{0DL}^{ikb} 
G_{0DR}^{\star ihq}
+\Gamma_{DL}^{kb} \Gamma_{DR}^{\star hq} G_{0DR}^{ihb} G_{0DL}^{\star ikq}  )
\sqrt{x_{\tilde{\chi}_i^0 \tilde{g}}}
F^{\prime}(x_{\tilde{d}_k \tilde{g}}, x_{\tilde{d}_h \tilde{g}},
x_{\tilde{\chi}_i^0 \tilde{g}})],
\nonumber \\
C_5^{\tilde g \tilde{\chi}^0}&=& \frac{\alpha_s \alpha_W}{4 m_{\tilde g}^2}
\sum_{h,k=1}^{6} \sum_{i=1}^9 [
( \Gamma_{DR}^{\star hq} \Gamma_{DL}^{\star kq} H_{0DR}^{ihb} H_{0DL}^{ikb}
-\frac{2}{3} \Gamma_{DR}^{ \star kq} \Gamma_{DL}^{\star hq} 
G_{0DR}^{ikb} G_{0DL}^{ihb}
-\frac{2}{3} \Gamma_{DR}^{kb} \Gamma_{DL}^{hb} G_{0DR}^{\star ikq} 
G_{0DL}^{\star ihq})
\nonumber \\
& &- \frac{2}{3} \Gamma_{DL}^{kb} \Gamma_{DL}^{\star hq} 
G_{0DR}^{ihb} G_{0DR}^{\star ikq}
- \frac{2}{3} \Gamma_{DR}^{hb} \Gamma_{DR}^{\star kq} G_{0DL}^{ikb} 
G_{0DL}^{\star ihq}  )
G^{\prime}(x_{\tilde{d}_k \tilde{g}},
x_{\tilde{d}_h \tilde{g}},
x_{\tilde{\chi}_i^0 \tilde{g}})
\nonumber \\
& &  + ( \frac{4}{3} \Gamma_{DL}^{\star kq} \Gamma_{DR}^{hb} 
G_{0DL}^{ikb} G_{0DR}^{\star ihq}
+\frac{4}{3} \Gamma_{DR}^{\star kq} \Gamma_{DL}^{hb} G_{0DL}^{\star 
ihq} G_{0DR}^{ikb}
\nonumber \\
& & + \frac{2} {3} \Gamma_{DR}^{hb} \Gamma_{DR}^{\star kq} 
G_{0DL}^{ikb} G_{0DL}^{\star ihq}
+ \frac{2}{3} \Gamma_{DL}^{kb} \Gamma_{DL}^{\star hq} G_{0DR}^{ihb} 
G_{0DR}^{\star ikq}  )
\sqrt{x_{\tilde{\chi}_i^0 \tilde{g}}}
F^{\prime}(x_{\tilde{d}_k \tilde{g}}, x_{\tilde{d}_h \tilde{g}},
x_{\tilde{\chi}_i^0 \tilde{g}})] .
\end{eqnarray}

\subsection{Hadronic Matrix Elements}
We follow the notations and parameterizations of Ref. 
\cite{becirevic}. The hadronic
matrix elements in  the vacuum insertion approximation (VIA) 
\cite{VIA} are given by
\begin{eqnarray}
\left < \bar{B}_d^0| Q_1 | B_d^0\right >_{VIA} &=& \frac{2}{3} 
m_{B_d}^2 f_{B_d}^2,
\nonumber \\
\left < \bar{B}_d^0| Q_2 | B_d^0\right >_{VIA} &=& -\frac{5}{12} 
\left ( \frac{m_{B_d}}{m_b+m_d} \right )^2
m_{B_d}^2 f_{B_d}^2,
\nonumber \\
\left < \bar{B}_d^0| Q_3 | B_d^0\right >_{VIA} &=& \frac{1}{12} \left 
( \frac{m_{B_d}}{m_b+m_d} \right )^2
m_{B_d}^2 f_{B_d}^2,
\nonumber \\
\left < \bar{B}_d^0| Q_4 | B_d^0\right >_{VIA} &=&  \left 
[\frac{1}{12}+\frac{1}{2} \left ( \frac{m_{B_d}}{m_b+m_d} \right )^2 
\right ]
m_{B_d}^2 f_{B_d}^2,
\nonumber \\
\left < \bar{B}_d^0| Q_5 | B_d^0\right >_{VIA} &=& \left [ 
\frac{1}{4}+\frac{1}{6} \left ( \frac{m_{B_d}}{m_b+m_d} \right )^2 
\right ]
m_{B_d}^2 f_{B_d}^2,
\end{eqnarray}
where $m_{B_d}$, $m_b$ and $m_d$ are the masses of the $B_d$ meson, b 
and d quark respectively.
The expressions for $\tilde{Q}_{1-3}$ are same as those of $Q_{1-3}$.

To take into account renormalization effects, we define the $B$ parameters as
\begin{equation}
\left < \bar{B}_d^0| Q_i(\mu)  | B_d^0\right >_{VIA} = \left < \bar{B}_d^0
| Q_i | B_d^0\right >_{VIA} B_i(\mu),~~~i=1, \ldots 5
\end{equation}
  where the numerical values of the renormalization functions and masses at the
$m_b$ scale are
\begin{eqnarray}
m_b(m_b)=4.6 ~\mathrm{GeV}, & & m_d(m_b)=5.4 ~\mathrm{MeV},
\nonumber \\
B_1(m_b)=0.87(4)^{+5}_{-4}, & & B_2(m_b)=0.82(3)(4),
  \nonumber \\
B_3(m_b)=1.02(6)(9), & & B_4(m_b)=1.16(3)^{+5}_{-7},
\nonumber \\
B_5(m_b)=1.91(4)^{+22}_{-7}.
\end{eqnarray}
The coefficients at the scale of $m_b$ are given by
\begin{equation}
C_r (m_b)=\sum_i \sum_s (b_i^{(r,s)}+ \eta c_i^{(r,s)} ) \eta^{a_i} C_s (M),
\end{equation}
where $\eta=\alpha_s(M)/\alpha_s(m_b)$ and we have chosen 
$M=(m_{\tilde{g}}+m_{\tilde{q}})/2$.
The numerical coefficients $a_i, b_i^{(r,s)}, c_i^{(r,s)}$ can be 
found in Ref. \cite{becirevic}.

Putting all the above together, we can calculate the mass difference 
$\Delta m_{B_d}$ and
the  CP asymmetry $a_{J/\psi K_s}$.  The off-diagonal element of the 
$B_d$ mass matrix
can be written as
\begin{equation}
M_{12}(B_d)=\frac{\left < B_d^0| H_{eff}^{\Delta B=2} | 
\bar{B}_d^0\right >}{2m_{B_d}},
\end{equation}
We define
\begin{eqnarray}
\Delta m_{B_d} &=& 2 |M_{12}(B_d)|,
\nonumber \\
a_{J/\psi K_s} &=& \sin 2 \beta_{eff},
\end{eqnarray}
where $ 2 \beta_{eff}=\mathrm{arg} M_{12}(B_d)$.

\section{Numerical Analysis}

We are interested in analyzing the case in which the supersymmetric
partners have masses around the weak scale, so we will assume relatively light
superpartner masses. All trilinear scalar couplings in the soft supersymmetry
breaking Lagrangian are assumed to be universal: $A_{ij}=A \delta_{ij}$ and
$\mu_{ij}=\mu \delta_{ij}$. We fix $A$ to be $100$ GeV, $\mu=200$ GeV and
$\tan \beta =5$ throughout the analysis.

When supersymmetry is softly broken, there is no reason to expect
that the soft parameters would be flavor blind, or that they would 
violate flavor
in the same way as in the SM.
The unconstrained LRSUSY model, similar to the unconstrained MSSM, 
allows for new sources of
flavor violation among generations. In the process $B \rightarrow X_s \gamma$
\cite{fn} and $B \rightarrow X_s l^+ l^-$ \cite{bsl1l2} we allowed 
for flavor violations
between the second and third families in  the down squark mass matrix only.
Here we consider the effects of flavor violations between the first and third
generation in both the up and down squark mass matrix.

We parameterize all the unknown soft breaking parameters coming mostly from the
scalar mass matrices using the mass insertion approximation \cite{MI}.
In this framework we choose a basis for fermion and sfermion states
in which all the couplings of these particles to neutral gauginos are flavor
diagonal. Flavor changes in the squark sector arise from the 
non-diagonality of the squark propagators.
The normalized flavor mixing parameters used are
\begin{eqnarray}
\label{massins}
\delta_{q,LL,ij}&=&\frac{(m^2_{q,LL})_{ij}}{m_0^2},~~~
\delta_{q,RR,ij}=\frac{(m^2_{q,RR})_{ij}}{m_0^2}, \nonumber \\
\delta_{q,LR,ij}&=&\frac{(m^2_{q,LR})_{ij}}{m_0^2},~~~
\delta_{q,RL,ij}=\frac{(m^2_{q,RL})_{ij}}{m_0^2},
\end{eqnarray}
where $m_0^2$ is the average squark mass and $(m^2_{q,AB})_{ij}$ are
the off-diagonal elements which mix squark flavors for both left- and
right- handed squarks with $q=u,d$, and
$A,B=L,R$. We diagonalize squark mass matrices numerically, which is valid
even  when the parameters are not perturbative.

We keep our analysis general, but to show our results, we select only 
one possible source of flavor
violation in the squark sector at a time, and assume the others 
vanish. All diagonal entries
in the squark mass matrix are set equal and we study the mass 
difference $\Delta
m_{B_d}$  and the CP asymmetry $a_{J/\psi K_s}$ as a function of the relevant
off-diagonal element.

We start by studying the constraints set by the experimental value of 
$\Delta m_{B_d}$.
We treat the real and imaginary parts of mass insertions as 
independent  parameters. As
discussed, upper bounds on various mass insertions are obtained by 
requiring  that the
contribution of each mass insertion saturate the experimental central 
value $\Delta
m_{B_d} <  0.489~ ps^{-1}$.

\begin{figure}
\centerline{ \epsfysize 4.0in \rotatebox{270}{\epsfbox{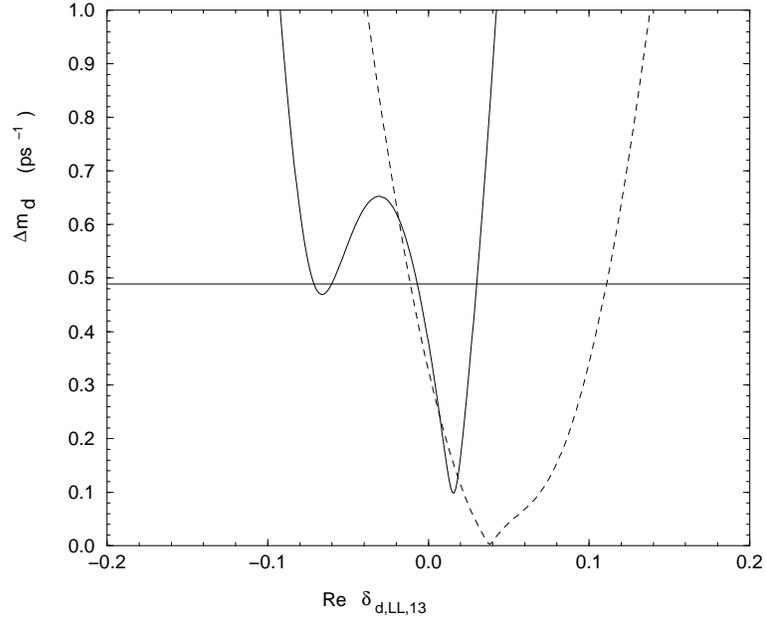}} }
\caption{$\Delta m_{B_d}$ as a function of $Re ~\delta_{d, LL, 13}$
with other mass insertion terms switched off. The solid(dashed)
line corresponds to $m_0/m_{\tilde{g}}=200/200 ~(200/400)$ GeV.
The straight line represents the experimental central value. }
\protect \label{deltamredLL13}
\end{figure}

\begin{figure}
\centerline{ \epsfysize 4.0in \rotatebox{270}{\epsfbox{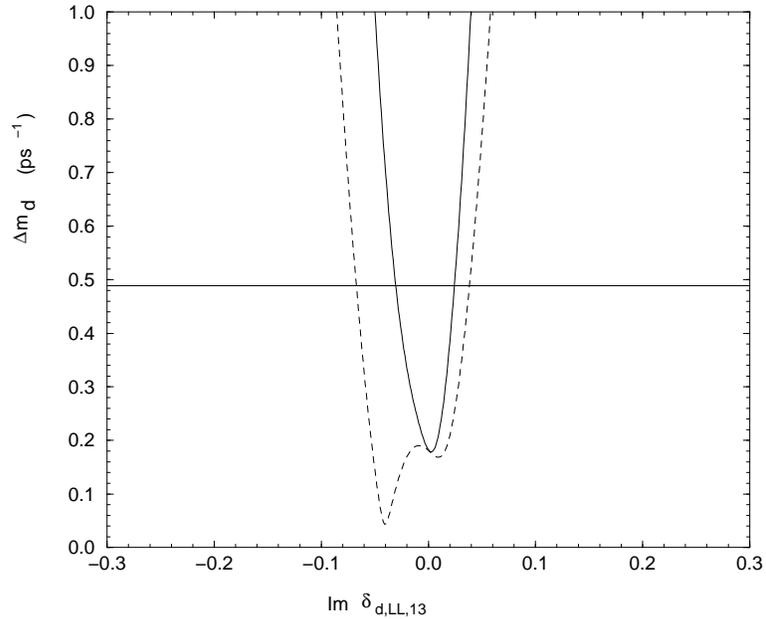}} }
\caption{$\Delta m_{B_d}$ as a function of $Im~ \delta_{d, LL, 13}$
with other mass insertion terms switched off. The solid(dashed) line 
corresponds to
$m_0/m_{\tilde{g}}=200/200~ (200/400)$ GeV.
The straight line represents the experimental central value.   }
\protect \label{deltamimdLL13}
\end{figure}

In Fig. \ref{deltamredLL13}, we show $\Delta m_{B_d}$ as a
function of $Re ~ \delta_{d, LL,13}$. Here $M_L=M_R=500$ GeV.
As the down squarks contribute in graphs
with exchange of gluinos, we give two values to the mass ratio 
$m_0/m_{\tilde{g}}$.
The curves are not completely symmetric around the zero point. For 
$m_0/m_{\tilde g}=200/200$
GeV, the upper bound of $|Re ~ \delta_{d, LL,13}|$ is found to be about $0.07$,
while for $m_0/m_{\tilde g}=200/400$ GeV, it is $0.11$. The bounds 
found here are
compatible  with the results of Ref. \cite{gk}.
Although we have chosen $\delta_{d, LL,13}$ as representative, very similar
constraints are obtained for $\delta_{d, RR, 13}$.

Even though $\Delta m_{B_d}$ is a CP conserving quantity, squark mass 
matrices can be
complex. The imaginary parts of the squark mass mixing give rise to 
the CP asymmetry
$a_{J/\Psi K_s}$. We could constrain the imginary parts from either 
the asymmetry, or
$\Delta m_{B_d}$ and we shall do both. We present the restrictions 
coming from  $\Delta
m_{B_d}$ first, and
  $a_{J/\Psi K_s}$ at the end of this section. In Fig. 
\ref{deltamimdLL13}, we show
$\Delta m_{B_d}$ as a function of $Im ~ \delta_{d, LL,13}$.
For $m_0/m_{\tilde g}=200/200$ GeV, the upper bound of $|Im ~ 
\delta_{d, LL,13}|$ is found to
be about $0.03$, while for  $m_0/m_{\tilde g}=200/400$ GeV, it is 
$6.5 \times 10^{-2}$.
Again, the constraints obtained for $Im ~\delta_{d, RR,13}$ are
very similar.

\begin{figure}
\centerline{ \epsfysize 4.0in \rotatebox{270}{\epsfbox{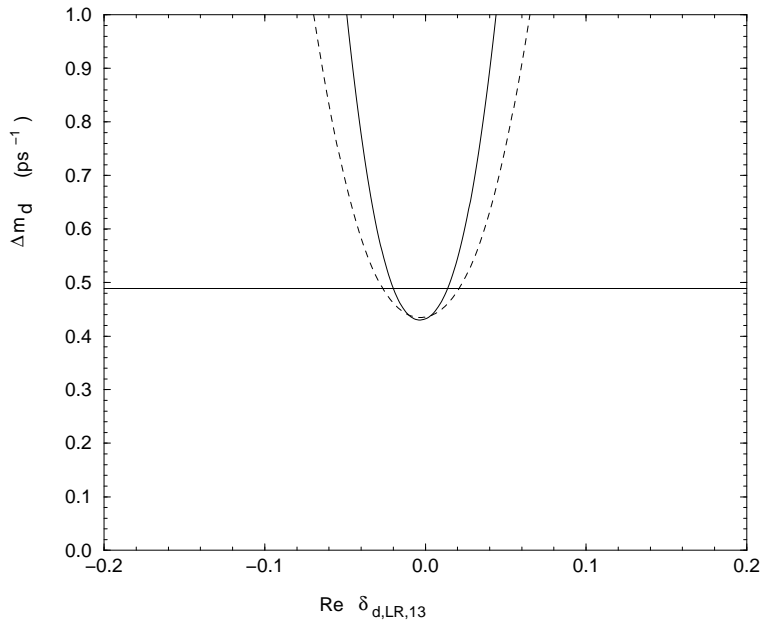}} }
\caption{$\Delta m_{B_d}$ as a function of $Re~ \delta_{d, LR, 13}$
with other mass insertion terms switched off. The solid(dashed)
line corresponds to $m_0/m_{\tilde{g}}=200/200~(200/400)$ GeV.
The straight line represents the experimental central value.  }
\protect \label{deltamredLR13}
\end{figure}

\begin{figure}
\centerline{ \epsfysize 4.0in \rotatebox{270}{\epsfbox{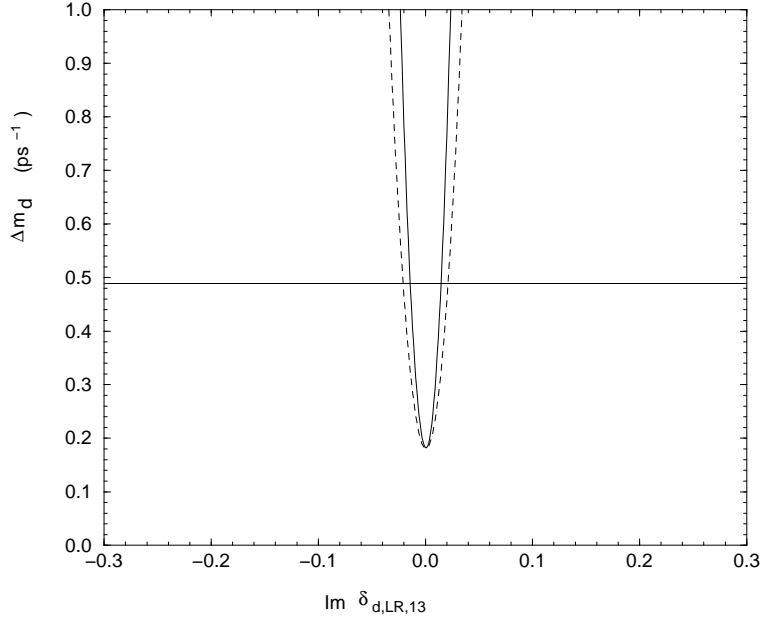}} }
\caption{$\Delta m_{B_d}$ as a function of $Im ~\delta_{d, LR, 13}$
with other mass insertion terms switched off. The solid(dashed) line 
corresponds to
$m_0/m_{\tilde{g}}=200/200~ (200/400)$ GeV.
The straight line represents the experimental central value.   }
\protect \label{deltamimdLR13}
\end{figure}

We proceed with an analysis of the chirality flipping flavor mixing 
parameters. In Fig.
\ref{deltamredLR13}, we show
$\Delta m_{B_d}$ as a function of $Re ~ \delta_{d, LR,13}$.
For $m_0/m_{\tilde g}=200/200$ GeV, the upper bound
of $|Re ~\delta_{d, LR,13}|$ is found to be about $2.5 \times 10^{-2}$,
while for $m_0/m_{\tilde g}=200/400$ GeV, it is $3.4 \times 10^{-2}$.
Similar constraints are
obtained for $\delta_{d, RL, 13}$. In Fig. \ref{deltamimdLR13} we show the
corresponding variation of $\Delta m_{B_d}$ with  $Im ~\delta_{d, LR,13}$.
The range of the
chirality flipping parameter $Im ~\delta_{d, LR,13}$ is more 
restrictive than the
chirality conserving $Im ~\delta_{d, LL,13}$ (both for the real and 
imaginary parts).

\begin{figure}
\centerline{ \epsfysize 4.0in \rotatebox{270}{\epsfbox{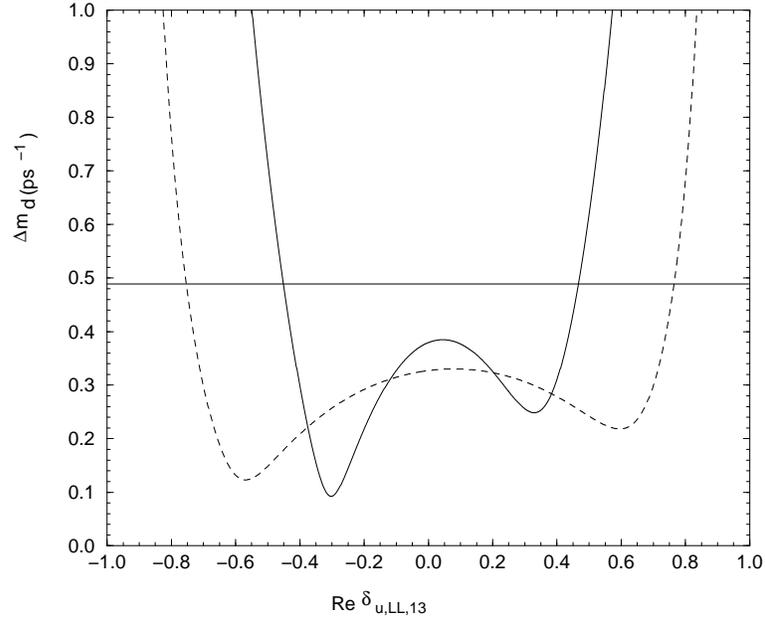}} }
\caption{$\Delta m_{B_d}$ as a function of $Re ~\delta_{u, LL, 13}$
with other mass insertion terms switched off. The solid(dashed) line 
corresponds to
$M_L=M_R=500 ~(1000)$ GeV. The straight line represents the 
experimental central value. }
\protect \label{deltamreuLL13}
\end{figure}

\begin{figure}
\centerline{ \epsfysize 4.0in \rotatebox{270}{\epsfbox{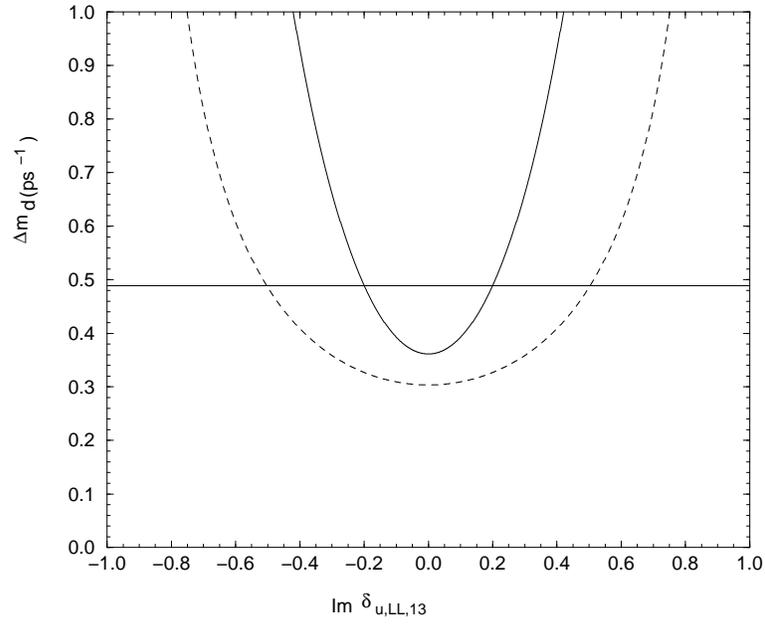}} }
\caption{$\Delta m_{B_d}$ as a function of $Im~ \delta_{u, LL, 13}$
with other mass insertion terms switched off. The solid(dashed) line 
corresponds to
$M_L=M_R=500 ~(1000)$ GeV. The straight line represents the 
experimental central value.  }
\protect \label{deltamimuLL13}
\end{figure}

If we turn off the down squark mass insertion, the $B^0_d-{\bar 
B}^0_d$ mass mixing
will be dominated by up squark mass insertions coming from diagrams 
with charginos in
the loop. We analyze these restrictions next.
Here we fix $m_0/m_{\tilde g}=200/200$ GeV and vary $M_L$ and $M_R$, 
equivalently to changing the masses
of charginos.  In Fig. \ref{deltamreuLL13}, we show the
dependence of $\Delta m_{B_d}$ on $Re ~ \delta_{u, LL,13}$.
For $M_L=M_R=500$ GeV, the lightest chargino masses are 
$m_{\chi^+_1}= 200$~GeV,
$m_{\chi^+_2}= 248$~GeV, and the upper bound of
$|Re ~ \delta_{u,LL,13}|$ is found to be  about $0.45$, while for 
$M_L=M_R=1000$ GeV,
the lightest chargino masses are
$m_{\chi^+_1}= 200$~GeV  and $m_{\chi^+_2}= 252$~GeV, and the bound 
is $0.75$.  Allowing
for imaginary parts of the flavor mixing only,  we show
$\Delta m_{B_d}$ as a function of $Im ~ \delta_{u, LL,13}$ in Fig. 
\ref{deltamimuLL13}.
For $M_L=M_R=500$ GeV, the upper
bound of
$|Im ~ \delta_{u, LL,13}|$ is found to be about $0.20$,
and for $M_L=M_R=1000$ GeV, it is $0.50$. Taking
$M_L=M_R$ is a simplification; the results depend on the chargino 
masses and cannot
distinguish separate values for the left or right gaugino masses. 
Very similar curves
are obtained for
$M_L \neq M_R$.

\begin{figure}
\centerline{ \epsfysize 4.0in \rotatebox{270}{\epsfbox{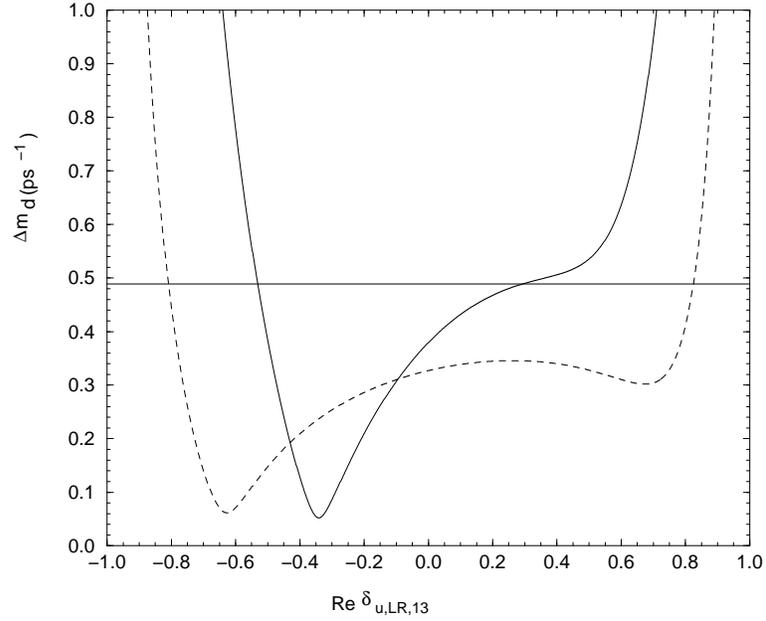}} }
\caption{$\Delta m_{B_d}$ as a function of $Re ~\delta_{u, LR, 13}$
with other mass insertion terms switched off. The solid(dashed) line 
corresponds to
$M_L=M_R=500 ~(1000)$ GeV. The straight line represents the 
experimental central value.  }
\protect \label{deltamreuLR13}
\end{figure}

\begin{figure}
\centerline{ \epsfysize 4.0in \rotatebox{270}{\epsfbox{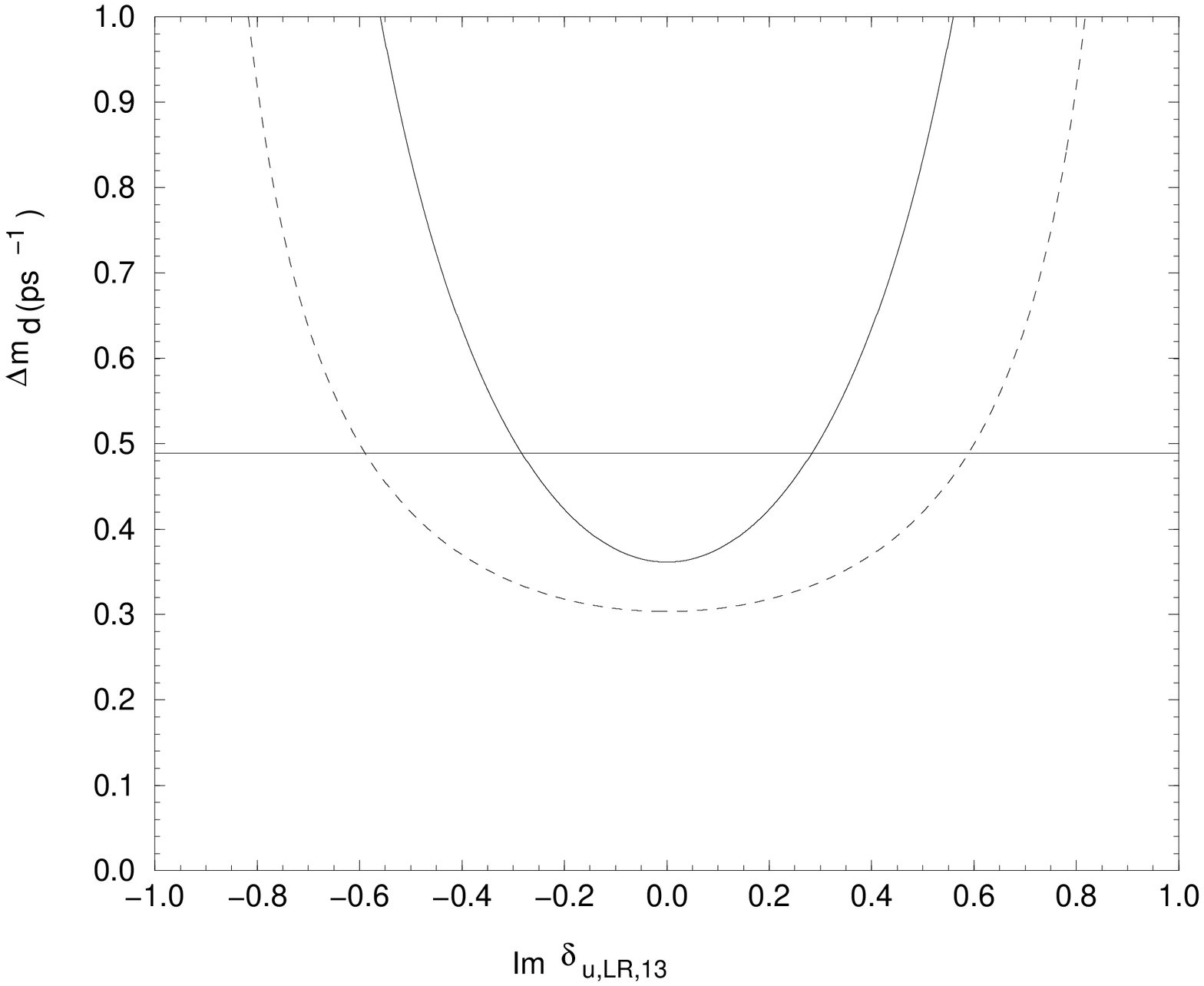}} }
\caption{$\Delta m_{B_d}$ as a function of $Im ~\delta_{u, LR, 13}$
with other mass insertion terms switched off. The solid(dashed) line 
corresponds to
$M_L=M_R=500 ~(1000)$ GeV. The straight line represents the 
experimental central
value.   }
\protect \label{deltamimuLR13}
\end{figure}

In Fig. \ref{deltamreuLR13}, we show $\Delta m_{B_d}$ as a function
of $Re ~ \delta_{u, LR,13}$, when only flavor violating,
chiralty flipping mass insertions in the up squark sector are non-zero.
For $M_L=M_R=500$ GeV, the upper bound of
$|Re ~ \delta_{u, LR,13}|$ is found to be about $0.53$;
while for $M_L=M_R=1000$ GeV, it is $0.81$.
And finally in Fig.
\ref{deltamimuLR13}, we show $\Delta m_{B_d}$ as a function of $Im ~ 
\delta_{u, LR,13}$.
For $M_L=M_R=500$ GeV, the upper bound of $|Im ~ \delta_{u, LR,13}|$ 
is found to
be about $0.28$; while for $M_L=M_R=1$ TeV, it is $0.59$.

We note that the conditions on the mass insertion in the up squark 
sector are much
less restrictive. This is due to the fact that these restrictions come from the
chargino contributions, which are smaller that the combined gluino, 
gluino-neutralino
and neutralino contribution. For the down squark mass insertions, the chirality
flipping mass insertions are more restricted than the chirality
conserving parts; while for the up squark mass insertions,  the chirality
conserving mass insertions are slightly more restricted than the 
chirality flipping parts.

\begin{figure}
\centerline{ \epsfysize 4.0in \rotatebox{270}{\epsfbox{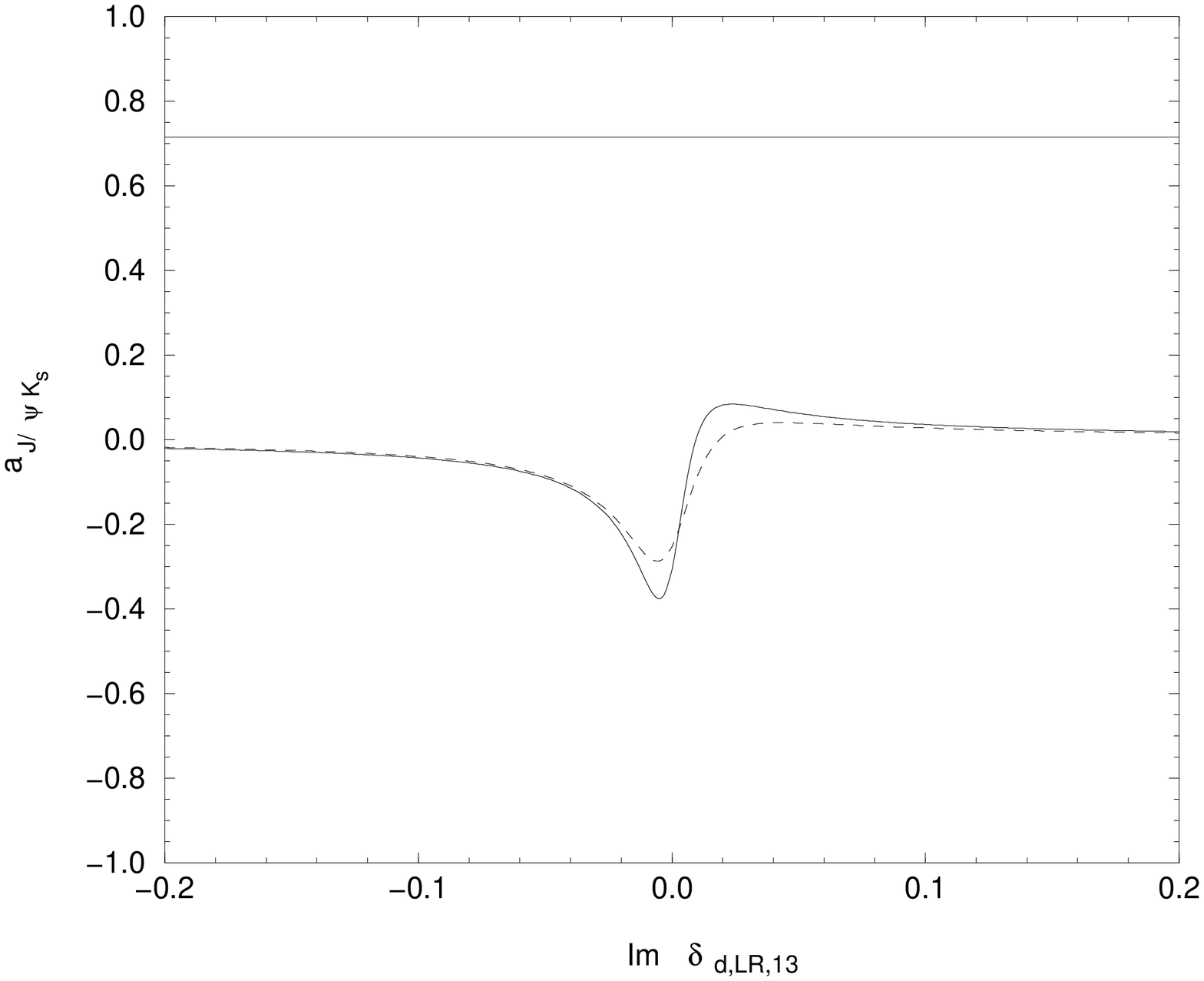}} }
\caption{$a_{J/\psi K_s}$ as a function of $Im ~\delta_{d, LR, 13}$
with other mass insertion terms switched off. The solid(dashed) line 
corresponds to
$m_0/m_{\tilde{g}}=200/200~ (200/400)$ GeV.
The straight line represents the experimental central value.    }
\protect \label{aJpsiKsimdLR13}
\end{figure}

\begin{figure}
\centerline{ \epsfysize 4.0in \rotatebox{270}{\epsfbox{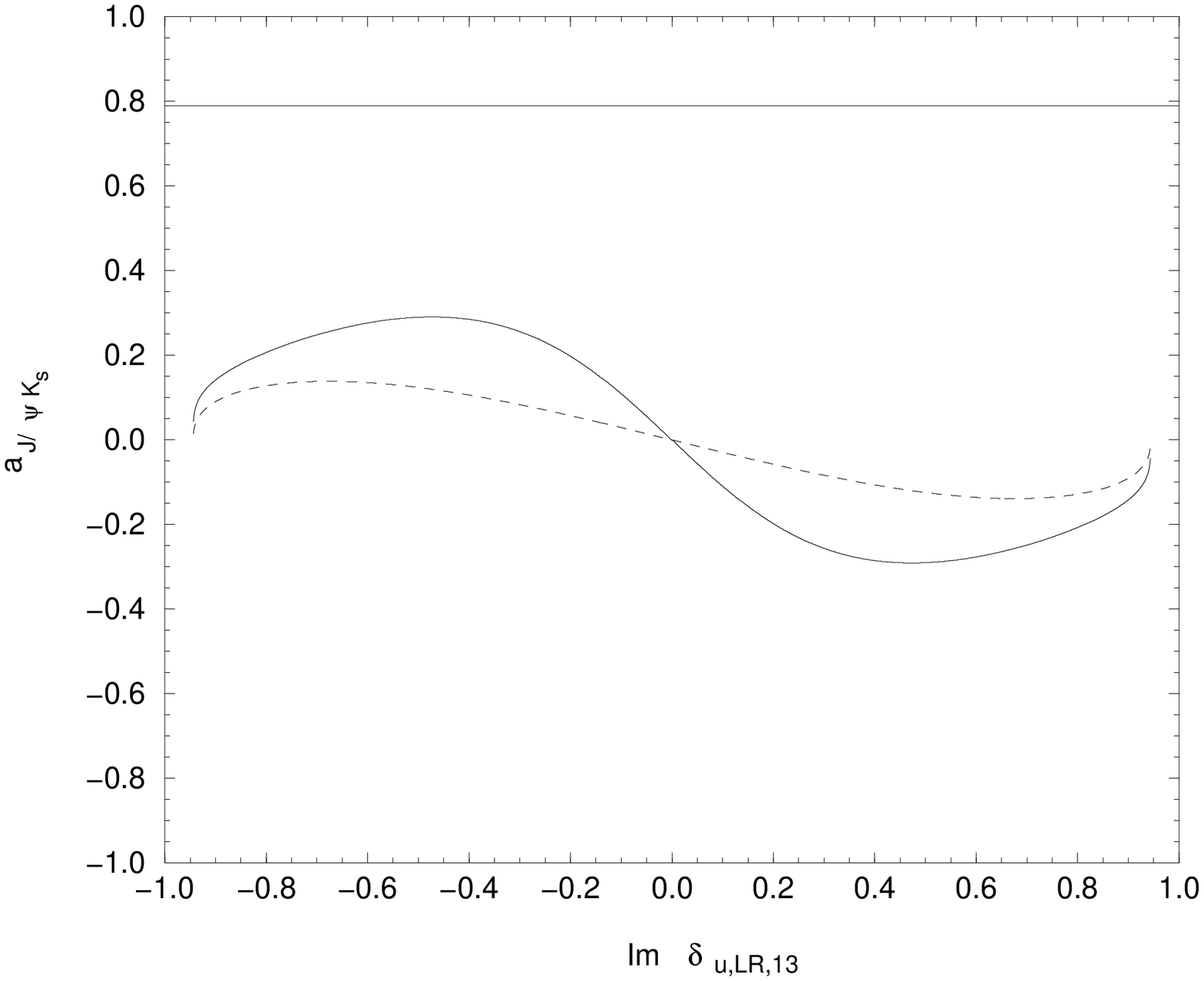}} }
\caption{$a_{J/\psi K_s}$ as a function of $Im~ \delta_{u, LR, 13}$
with other mass insertion terms switched off. The solid(dashed) line 
corresponds to
$M_L=M_R=500 ~(1000)$ GeV. The straight line represents the 
experimental central value.  }
\protect \label{aJpsiKsimuLR13}
\end{figure}

We now turn to study the CP asymmetry $a_{J/\psi K_s}$. The imaginary 
parts of various
mass insertions provide additional complex phases to violate CP. 
Switching off the SM CKM
phase, we try to constrain the imaginary parts by requiring that 
contributions of the
additional SUSY phases to $\sin 2 \beta$ do not exceed its  world 
average central value
$0.79$. In Fig.  \ref{aJpsiKsimdLR13} and \ref{aJpsiKsimuLR13}, we 
show $a_{J/\psi K_s}$
as a function of $Im ~ \delta_{d,LR, 13}$ and $Im ~ \delta_{u,LR, 
13}$. It is found
that the LRSUSY contribution only to $a_{J/\psi K_s}$ is in general 
small and below the
central value. The asymmetry even becomes
negative for some parameter values. Not knowing the relative phases of this and
the SM contribution, we cannot make a firm prediction. Therefore, unlike models
with non-universal soft terms
\cite{gk}, we do not obtain constraints
on the imaginary parts of mass inserions from
$a_{J/\psi K_s}$ in this model. But depending on the unknown phase, 
the prediction for
the asymmetry in LRSUSY could be different than in the SM, especially 
in the region of
$|Im~ \delta_{d, LR, 13}| \leq 10^{-2}$.

We also note that in previous figures $Im ~ \delta_{u, LR(RL),13}$, 
$Im ~ \delta_{d,
LR(RL),13}$ are allowed to be quite large. These appear to
contradict experimental bounds coming from the electric dipole 
moments. The most severe
restriction comes from the measured electric dipole moment of the Mercury atom:
\begin{equation}
d_{Hg}<2.1 \times 10^{-28} e~cm
\end{equation}
which would require $Im ~ \delta_{u,d LR(RL),11} \le 10^{-7}-10^{-8}$ 
\cite{bgkt}.
In LRSUSY parity symmetry forces the Yukawa couplings and the trilinear
$A$ terms to be Hermitean above the $\Lambda_R$ scale \cite{mr}. This 
insures that the
diagonal elements of $A_{ij}$ and $Y_{ij}$ are real in any basis 
\cite{abkl}. However, it
appears that hermiticity alone cannot prevent large contributions 
being induced into
diagonal elements by large off-diagonal $Im~ \delta_{u, LL, 13}$ or 
$Im ~\delta_{u, LR,
13}$, which would give too large values for $d_n$. Analyses performed elsewhere
in the literature \cite{sk} suggest that some other phenomenology, such as
non-universality of scalar masses, has to be invoked to explain both the
large CP violation observed in the B and K sectors, and the smallness 
of the EDM-s.
In this analysis we prefer to set the constraints coming from the CP
violation in the $B_d^0-{\bar B}_d^0$ sector independent of other 
constraints.  It
is known that EDM-s can be suppressed by heavy sfermions, cancellations 
between different
dipole contributions, or small CP phases. Given the array of 
possibilities, we do not
choose a particular additional condition on the LRSUSY. But we 
emphasize that, unless
additional parameters are introduced, or constraints relaxed, the 
imaginary part of the
mass insertions would have to be much smaller than that imposed by 
$B^0_d-{\bar B}^0_d$
mixing alone.

\section{Conclusions}

We have studied $B_d^0- \bar{B}^0_d$ mixing and the CP asymmetry 
$a_{J/\psi K_s}$ in the fully
left-right supersymmetric model. Explicit expressions for all the
chargino, gluino, gluino-neutralino and neutralino amplitudes involved in
the process are given. We obtain conservative constraints on the 
various squark mass
splittings  by imposing that the supersymmetric contributions do not exceed the
central value of the measured $\Delta m_{B_d}$.

Throughout our analysis, we found the gluino and chargino to
dominate the supersymmetric contributions; the neutralino and gluino-neutralino
contributions are smaller. If the dominant flavor mixing comes from 
the up-squark
sector, the chargino contribution dominates for large $\delta_{u,13}$ 
and quickly
saturates the LRSUSY contribution. If the only source of flavor 
mixing arises from the
down-squark region, the gluino contribution dominates for most the 
parameter space.
If both sources of flavor
violation are present, the gluino contribution will saturate the 
experimental bound
faster, justifying most of the previous analyses which looked at the gluino
contributions only. However, we stress the importance of analyzing 
both the up squark
and down squark sources of flavor violation for a complete picture of 
the $B_d^0-
\bar{B}^0_d$ mixings. The asymmetry $a_{J/\psi K_s}$ is within the 
allowed range for all
values of $Im ~ \delta_{u, 13}$ and $Im ~ \delta_{d, 13}$, though in 
LRSUSY the new
sources of flavor and CP violation could enhance this parameter over the
SM contribution in the region where $|Im~ \delta_{d, LR, 13}| \leq 10^{-2}$.

Comparisons with other models show that some general features are similar: the
chirality flipping mass insertions are more restricted than the 
chirality conserving
  ones and the down squark mixings are more restricted than the up 
squark mixings
\cite{becirevic}. A more general conclusion
escapes us because a comparable comprehensive analysis of mass 
insertions in the
unconstrained MSSM does not exist for $B_d^0- \bar{B}^0_d$ mixing.

FCNC and CP violating phenomena in B physics are promising
candidates for indirect SUSY signals and complementary to the direct 
searches. Thus
efforts to improve the theoretical precision in various SUSY 
scenarios are necessary.
The present analysis is useful in restricting all mass insertions, 
taken as independent
parameters, in a supersymmetric model with left-right symmetry. The 
dependence on the
details of the model, beyond the requirement of left-right symmetry, 
(such as the
triplet Higgs structure) is minimal and negligible. This analysis 
restricts further the
FCNC CP conserving and CP violating parameters in the squark sector 
of LRSUSY, and
provides complementary information to the one extracted from $B 
\rightarrow X_s \gamma$
and $B_s \rightarrow X_s l_1 l_2$.

\vskip0.2in

\noindent {\bf Acknowledgements}

This work was funded in part by NSERC of Canada (SAP0105354).

\def\oldprd#1#2#3{{\rm Phys. ~Rev. ~}{\bf D#1}, #3 (19#2)}
\def\newprd#1#2#3{{\rm Phys. ~Rev. ~}{\bf D#1}, #3 (20#2)}
\def\plb#1#2#3{{\rm Phys. ~Lett. ~}{\bf B#1}, #3 (#2)}
\def\newplb#1#2#3{{\rm Phys. ~Lett. ~}{\bf B#1}, #3 (20#2)}
\def\npb#1#2#3{{\rm Nucl. ~Phys. ~}{\bf B#1}, #3 (19#2)}
\def\newnpb#1#2#3{{\rm Nucl. ~Phys. ~}{\bf B#1}, #3 (20#2)}
\def\prl#1#2#3{{\rm Phys. ~Rev. ~Lett. ~}{\bf #1}, #3 (19#2)}
\def\prl20#1#2#3{{\rm Phys. ~Rev. ~Lett. ~}{\bf #1}, #3 (20#2)}
\def\rep19#1#2#3{{\rm Phys. ~Rep. ~}{\bf #1}, #3 (19#2)}
\def\rep20#1#2#3{{\rm Phys. ~Rep. ~}{\bf #1}, #3 (20#2)}
\def\epjc#1#2#3{{\rm Eur. ~Phys. J.~}{\bf C#1}, #3 (#2)}

\bibliographystyle{unsrt}

\end{document}